\documentstyle[prl,aps,epsfig]{revtex}

\begin{document}
\draft

\topmargin = .000001in

\twocolumn[\hsize\textwidth\columnwidth\hsize\csname @twocolumnfalse\endcsname
\title{Microscopic Structure of a Vortex Line in a Dilute Superfluid Fermi Gas}

\author{N.~Nygaard$^{1,2}$, G.~M.~Bruun$^3$, C.~W.~Clark$^1$, and
D.~L.~Feder$^4$}  
\address{$^1$Electron and Optical Physics Division, National
Institute of Standards and  Technology, Gaithersburg, Maryland
20899-8410}
\address{$^2$Chemical Physics Program, University of Maryland, College Park, 
Maryland}
\address{$^3$Niels Bohr Institute, Blegdamsvej 17, 2100 Copenhagen, Denmark}
\address{$^4$Department of Physics and Astronomy, University of Calgary, 
Calgary, Alberta, Canada T2N 1N4}
\date{\today}
\maketitle

\begin{abstract} 

The microscopic properties of a single vortex in a dilute superfluid Fermi gas 
at zero temperature are examined within the framework of self-consistent 
Bogoliubov-de Gennes theory. Using only physical parameters as input, we study 
the pair potential, the density, the energy, and the current distribution.
Comparison of the numerical results with analytical expressions clearly 
indicates that the energy of the vortex is governed by the zero-temperature BCS 
coherence length.

\end{abstract}

]


The trapping and cooling of dilute Fermi gases is an increasingly
active area of research within the field  
of ultracold atomic gases. Several experimental groups now trap and
cool alkali atoms with Fermi statistics reaching temperatures   
as low as $\sim0.2T_F$ with $T_F$ being the Fermi temperature of the
system~\cite{Experiments}. These gases are appealing to study due to
the large experimental control of their properties and the
microscopically well-understood two-body effective interaction between
the atoms. One of the main goals of the present experimental effort is  
to observe a phase transition to a superfluid state predicted to occur
below a certain critical temperature $T_c$~\cite{StoofBCS}. 
A fascinating prospect of such a superfluid is the formation
of quantized vortices. The combination of theoretical and experimental
studies of vortices in Bose-Einstein condensates (BECs) have produced a
number of beautiful results in recent years~\cite{Fetter}. For
fermions, the nature of vortices in different systems such as type-II
superconductors, superfluid $^3$He, and neutron stars is a classic problem 
with a vast literature~\cite{deGennes,He3,ElgaAstro}.  

One problem of fundamental importance is to calculate the energy of a
vortex for $T=0$. This energy is defined as the difference between the
energy of the superfluid state with the vortex present and without.
Dividing this value by the
angular momentum per particle in the vortex state gives the frequency at which 
the vortex becomes the thermodynamic ground state of a rotating system.
It is rather surprising that despite the large amount of work concerned with 
the structure of a vortex for a fermionic system, there is  
no clear result regarding this specific problem. This is in contrast
with the case of a dilute bosonic superfluid, where the
Gross-Pitaevskii equation allows an analytical calculation of the vortex
energy for $T=0$ if quantum fluctuations are
neglected~\cite{Gross}. The equivalent 
theory relevant for fermions, Ginzburg-Landau theory, is unfortunately only  
valid for $|T-T_c|/T_c\ll1$ making an analytical calculation of the
energy for $T=0$ more complicated. In this letter we present the first
{\it{ab-initio}} calculation of the vortex energy based on microscopic theory.

We consider in the following a two-component gas of
neutral fermionic atoms with mass $m$ in a cylinder of radius $R$. 
We take
the number of particles in each spin state $N_{\sigma}$ to be the
same as this is the optimum situation for superfluidity~\cite{StoofBCS}.
In the dilute limit, the interaction between the atoms in the two spin
states $\sigma=\uparrow,\downarrow$ can be well described by the
contact potential $g\delta({\mathbf{r}})$ where $g=4\pi\hbar^2 a/m$
and $a<0$ is the $s$-wave scattering length describing low energy
collisions between $\uparrow$ atoms and $\downarrow$ atoms. 
In the zero-temperature limit that we are treating here there are no
intra-component collisions~\cite{pwavethreshold}.
Recently, two papers have
calculated the $T=0$ vortex energy for a Fermi superfluid under these assumptions. 
Using phenomenological models,\ the energy of a unit circulation vortex was
estimated in Ref.~\cite{BruunViverit} to be 
\begin{equation}\label{vorten}
{\cal{E}}_{v}\simeq \frac{\pi\hbar^2n_{\sigma}}{2m} 
\ln\left(D\frac{R}{\xi_{\rm BCS}}\right),
\end{equation}
where $n_\sigma=\tilde{k}_F^3/6\pi^2$ is the density in each of the
two components, and
$\xi_{\rm BCS}=\hbar^2 \tilde{k}_{F}/\pi m\Delta_{0}$ is the BCS coherence length 
with $\Delta_0$ the bulk value of the superfluid gap; BCS theory predicts 
$\Delta_0=8{\mathrm{e}}^{-2}\tilde{\mu}\exp(-\pi/2\tilde{k}_F|a|)$
~\cite{Induced}. The effective Fermi momentum, $\tilde{k}_F$, is defined as
$\hbar^2\tilde{k}_{F}^2/2m=\mu-gn_\sigma\equiv\tilde{\mu}$, 
where $\mu$ is the chemical potential,
such that it includes the effect of
the Hartree mean-field. 
It was argued in Ref.~\cite{BruunViverit}
that a microscopic calculation for $T=0$ would yield $D$ to be a \emph{constant}
$\sim{\mathcal{O}}(1)$  independent of $\tilde{k}_F$ and $|a|$ 
since the characteristic length-scale of a vortex must be expected to
be ${\mathcal{O}}(\xi_{\rm BCS})$.
The value of $D$ depends on the phenomenological
model used: If the vortex is modeled as a cylinder of radius
$\xi_{\rm BCS}$ containing a normal stationary fluid,  
surrounded by a rotating superfluid 
one obtains $D=1.36$. We refer to this simple
model as the cylinder model. If Ginzburg-Landau theory is applied, we
get $D=1.65$.

This conclusion was however disputed in the work of
Ref.~\cite{ElgaroyVortex}. Here it was argued that the  
characteristic length scale of the vortex is much smaller than
$\xi_{\rm BCS}$ and the energy correspondingly higher. This is because the
structure of the vortex core is determined by the lowest lying vortex
states. These states are formed out of excitations around the Fermi
level with typical wavelengths $\sim\tilde{k}_F^{-1}$, 
and following the conclusions based on the analytical and
numerical solutions of the Bogoliubov-de Gennes (BdG)
equations~\cite{Kramer,Gygi} it was argued that the important length
scale of the core region is $\xi_1=4/\pi
\tilde{k}_F^2|a|\ll\xi_{\rm BCS}$ in the dilute regime~\cite{ElgaroyVortex}. 
Using $\xi_1$ as the size of the vortex core, a
calculation identical  to the one given in Ref.~\cite{BruunViverit} leads
to a vortex  energy given by Eq.~(\ref{vorten}) but with  
$D\simeq\xi_{\rm BCS}/\xi_1\gg 1$ in the dilute regime. Thus, the energy
was predicted to be significantly higher than  
what was estimated in Ref.~\cite{BruunViverit}. Note that $D$ is now
not a constant but depends on $\tilde{k}_F$ and $|a|$. 

It is presently not clear which of the two quite different predictions is
correct and thus what the energy of the vortex actually is. 
In order to settle this question, we now present a \emph{microscopic}
calculation of the vortex energy using the assumptions given above.
The BdG equations describing 
the superfluid state read~\cite{deGennes} 
\begin{eqnarray}
\left[\begin{array}{cc}H_0({\mathbf{r}})&\Delta({\mathbf{r}})\\
\Delta^*({\mathbf{r}})&-H_0({\mathbf{r}}) 
\end{array}
\right]\left[\begin{array}{c}u_\eta({\mathbf{r}})\\v_\eta({\mathbf{r}})\end{array}
\right]=
E_\eta\left[\begin{array}{c}u_\eta({\mathbf{r}})\\v_\eta({\mathbf{r}})\end{array}
\right],
\end{eqnarray}
with $H_0({\mathbf{r}})=-\hbar^2\nabla^2/2m-\mu+U({\mathbf{r}})$, and where
$U({\mathbf{r}})=gn_\sigma({\mathbf{r}})
=g\sum_{\eta}|v_{\eta}({\mathbf{r}})|^2$ 
is the Hartree field and
$\Delta({\mathbf{r}})=-\tilde{g}\sum_{\eta}u_{\eta}({\mathbf
r})v_{\eta}^*({\mathbf r})$ is the gap function. To
avoid having to introduce an arbitrary high energy cut-off in the
theory, we have used a zero range pseudo-potential scheme giving rise
to a regularized coupling constant $\tilde{g}$, when
calculating $\Delta({\mathbf{r}})$~\cite{BruunBCS}. The result is a
well-defined theory using only physical parameters as input. Further details
of this and the numerical techniques used to solve these equations will be given 
elsewhere. Once the
self-consistent solution is obtained for a given coupling strength and 
chemical potential, the energy is given by
$E=\langle\hat{H}-\mu\hat{N}\rangle$, where $\hat{H}$ is the Hamiltonian of 
the system~\cite{deGennes}, and $\hat{N}$ is the number operator. Defining
$E({\bf r})\equiv 2\sum_\eta|v_{\eta}({\mathbf{r}})|^2 E_\eta$, $E$ can be 
calculated using

\begin{equation}
E=-\int d^3r \ 
\left\{
E({\bf r})
+\frac{1}{\tilde{g}}\left[|U({\mathbf{r}})|^2+|\Delta({\mathbf{r}})|^2\right]\right\}.
\label{Energy}
\end{equation}

We now solve the BdG equations for the gas in a cylinder of height $L_z$ and radius
$R\gg\xi_{\rm BCS}$. Since $\xi_{\rm BCS}$ is a decreasing function 
of $\tilde{k}_F|a|$
we use two different cylinder sizes. For $\tilde{k}_F|a|\le 0.43$ 
$(\tilde{k}_F|a|\ge 0.43)$ we take
$R=44.1 \mu {\mathrm m}$ $(R=12.6 \mu {\mathrm m})$ and 
$L_z=12.5 \mu {\mathrm m}$ $(L_z=5 \mu {\mathrm m})$.
We find the lowest energy superfluid state of the 
system by setting $\Delta({\mathbf{r}})=\Delta(\rho,z)$ where $\rho$ is the
perpendicular distance from the axis of symmetry of the cylinder, and $z$ is
the axial coordinate. For a vortex state, we assume the form
$\Delta({\mathbf{r}})=\exp(-i\phi)\Delta(\rho,z)$ where $\phi$ is
the azimuthal angle around the cylinder axis. This corresponds to a vortex
line with unit circulation along the cylinder axis. 
In both cases $U({\mathbf{r}})=U(\rho,z)$. In the vortex-free case, the
cylindrical symmetry dictates that Cooper pairs form
between particles with angular momentum $\hbar \nu$ and $-\hbar \nu$ along
the cylinder axis whereas in the vortex case pair constituents have
angular momentum $\hbar \nu$ and $-\hbar (\nu+1)$~\cite{deGennes}. Once the
two solutions with and without the vortex   
are obtained, the energy per unit length of the vortex line can be
determined as ${\cal{E}}_{v}=(E_v-E_0)/L_z$ where $E_v$ denotes the
energy of the vortex state and $E_0$ the energy of the state without a
vortex, both obtained from Eq.~(\ref{Energy}).
Throughout the succeeding analysis we consider a gas of $^6$Li atoms
with $a=-955$ $ a_0$. The density in each
hyperfine state is chosen to be a few times $10^{13} \mathrm{cm}^{-3}$. These
parameters are appropriate for on-going experiments~\cite{JohnThomas}.
The numerical solution of the BdG equations is obtained within a
Bessel function discrete variable representation in $\rho$~\cite{FederDVR,Lemoine}, 
and periodic boundary conditions along the vortex axis.

\begin{figure}
\centering
\epsfig{file=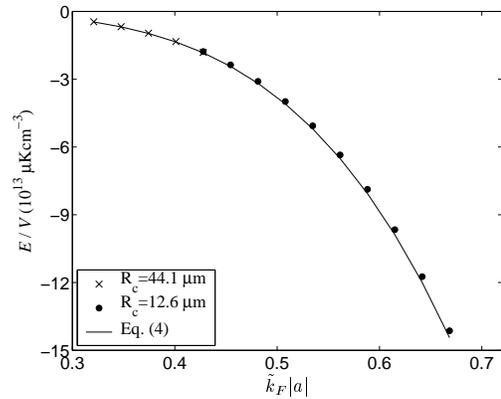,width=5.25cm,angle=-90}
\caption{The energy density of the superfluid gas.
Note that on this scale the energies of the system with and without a vortex
are indistinguishable, as the  
energy cost associated with vortex formation is much smaller
than the total energy.}
\label{Etot}
\end{figure}

In Fig.~\ref{Etot}, we plot the total energy density $E/V$  of the
superfluid gas, where $V$ is the volume of the cylinder, and $E$ is 
given by Eq.~(\ref{Energy}), as a function of  the effective
interaction strength $\tilde{k}_F|a|$. 
The $\times$'s and $\bullet$'s are obtained from a
self-consistent numerical solution of the BdG equations for two different 
values of $R$, whereas the line
is the ground state energy per unit volume of a bulk Fermi superfluid,  
\begin{equation}\label{Ebulk}
\frac{E_{bulk}}{V}=\frac{6}{5}n_\sigma\frac{\hbar^2\tilde{k}_F^2}{2m}-\mu2n_\sigma+
gn_\sigma^2-\frac{N(0)\Delta_{0}^2}{2}.  
\end{equation}
Here 
$N(0)=m\tilde{k}_F/2\pi^2\hbar^2$ is the density of states at the 
Fermi level.   
This expression is obtained by integrating analytically
Eq.~(\ref{Energy}) for a homogeneous gas where  
the $u({\mathbf{r}})$'s and $v({\mathbf{r}})$'s are simple plane wave states. The 
first three terms in  Eq.~(\ref{Ebulk}) give to the energy of a
homogeneous gas in the normal phase within the 
Hartree-Fock approximation and the last term is the condensation
energy. We see that there is good agreement
between our numerical results and the analytical formula. The slight
discrepancy is due to boundary effects at the edge of the cylinder,
where the density of particles vanishes. 
Note that we are at the limit of the weak coupling regime 
$\tilde{k}_F|a|\ll1$, appropriate for dilute gases. For the purposes of 
comparison with analytical results, however, it is important to calculate 
properties for the widest possible range of $\xi_{\rm BCS}$, subject to the 
condition $\xi_{\rm BCS}\ll R$ which ensures that the gap function can heal 
to its bulk 
value before becoming suppressed at the cylinder surface.

\begin{figure}
\centering
\epsfig{file=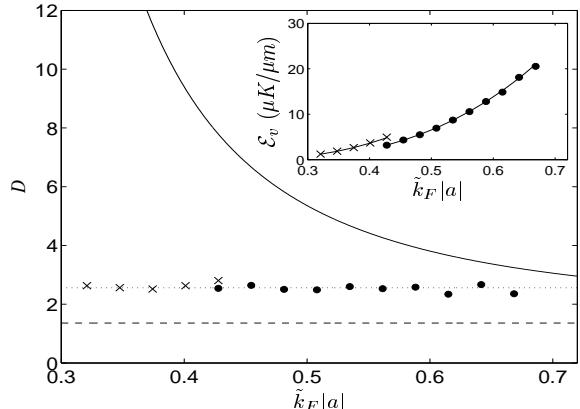,width=5.5cm,angle=-90}
\caption{The energy of the vortex in terms of the parameter $D$. The
dashed and solid lines correspond to the analytical predictions of Ref.~\protect\cite{BruunViverit} and~\protect\cite{ElgaroyVortex}, respectively, 
and the numerical results are indicated with $\bullet$'s ($R=12.6 \mu {\mathrm m}$)
and $\times$'s ($R=44.1 \mu{\mathrm m} $) with the average $\bar{D}$
represented by the dotted line. The inset depicts ${\cal{E}}_v$ with lines
giving the analytical prediction of Eq.~\ref{vorten} using $D=\bar{D}$.} 
\label{Denergy}
\end{figure}

In Fig.~\ref{Denergy}, we plot the numerically calculated energy of the vortex
${\cal{E}}_v$ for varying $\tilde{k}_F|a|$. To compare
with the analytical predictions, we parameterize  ${\cal{E}}_{v}$ by
the variable $D$ appearing in Eq.~(\ref{vorten}).
The dashed line corresponds to the prediction 
$D=1.36$~\cite{BruunViverit} and the solid line to
$D=\xi_{\rm BCS}/\xi_1$~\cite{ElgaroyVortex}. We see that the two
predictions for $D$ have a completely different dependence on
$\tilde{k}_F|a|$. The important conclusion is that the numerical
results confirm $D\sim{\mathcal{O}}(1)$ being a \emph{constant} 
independent of $\tilde{k}_F|a|$ in agreement with Ref.~\cite{BruunViverit}.
On the other hand, the 
prediction $D=\xi_{\rm BCS}/\xi_1$ yields a qualitatively incorrect
result. 
We note that the kink in ${\mathcal{E}}_v$
is due to $R/\xi_{\rm BCS}$ being different for the two cylinder sizes, whereas 
the spread in $D$ at $\tilde{k}_F|a|=0.43$ 
is indicative of our numerical accuracy. 
The numerical value $D\simeq 2.5$ is higher than the prediction of 
phenomenological models in Ref.~\cite{BruunViverit}. This is
as expected since these models only can yield the correct order of
magnitude of the constant inside the logarithm. Thus, the
length scale determining the energy of the vortex is
$\sim\xi_{\rm BCS}$ and not $\xi_1$. 

\begin{figure}
\centering
\epsfig{file=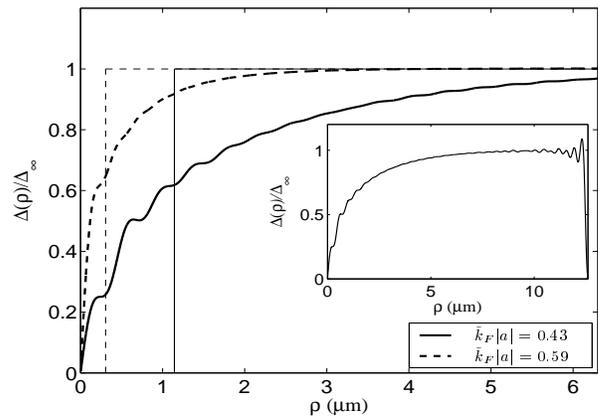,width=5.5cm,angle=-90}
\caption{The vortex profile $\Delta(\rho)/\Delta_\infty$ for two
values of $\tilde{k}_F|a|$. 
The thick solid line corresponds to 25,000
atoms per hyperfine state and a transition temperature of
$0.045\mu\mathrm{K}$ while the thick dashed curve is for $N_{\sigma}=66,500$
giving $T_c=0.23\mu\mathrm{K}$. For both curves $R=12.6 \mu {\mathrm m}$.
The thin solid and 
dashed lines depict the cylinder model of the vortex with radius
$\xi_2=0.42\xi_{\rm BCS}$ for the two $\tilde{k}_F|a|$ values. The inset shows
the full $\tilde{k}_F|a|=0.43$ solution.}
\label{VortexProfile}
\end{figure}

To examine this in more detail, we
plot in  Fig.~\ref{VortexProfile} the 
numerically calculated profile of a  vortex for two
representative values of $\tilde{k}_F|a|$.
Close to the vortex core, only the lowest-energy (bound) states contribute to 
the order parameter; these give rise to the observed Friedel 
oscillations, which have a wavelength on the order of
$\tilde{k}_F^{-1}$. We see that the length scale defined as 
$\xi_1=\lim_{\rho\rightarrow
0}[\Delta(\rho,z)/\rho\Delta_\infty]^{-1}$  
giving the slope of $\Delta({\mathbf{r}})$ at the vortex core actually
is much smaller than $\xi_{\rm BCS}$ as predicted in
Ref.~\cite{ElgaroyVortex,Kramer}. Here, $\Delta_\infty$ is the   
value of $|\Delta({\mathbf{r}})|$ far away from the vortex
core with $\Delta_\infty\simeq\Delta_0$ as expected. 
However, as the distance $\rho$ from the  
vortex core increases, the slope decreases and $\Delta({\mathbf{r}})$
reaches the value $\Delta_\infty$ on a length scale $\sim\xi_{\rm BCS}$
and not $\xi_1$. To quantify this, we use the cylinder model of the
vortex with a vortex radius $\xi_2=x\xi_{\rm BCS}$ 
 to calculate ${\cal{E}}_{v}$. This yields 
Eq.~(\ref{vorten}) but now with $D=(1.36)^{x^2}/x$. The equation
$D=2.5$ then gives $x=0.42$.
Thus $\xi_{2}=0.42\xi_{\rm BCS}$ is the length scale determining  the
energy of the vortex. Again, it should be emphasized that
$x\simeq0.42$ is a \emph{constant} over the large range of $\xi_{\rm BCS}$  
used in the calculations thereby verifying that indeed $\xi_{\rm BCS}$
determines the length-scale relevant for the energy as discussed in
\onlinecite{BruunViverit}. The cylinder model of $\Delta(\rho)$ with the 
correct radius $\xi_2=0.42\xi_{\rm BCS}$ is plotted in 
Fig.~\ref{VortexProfile}.

\begin{figure}[tbh]
\centering
\epsfig{file=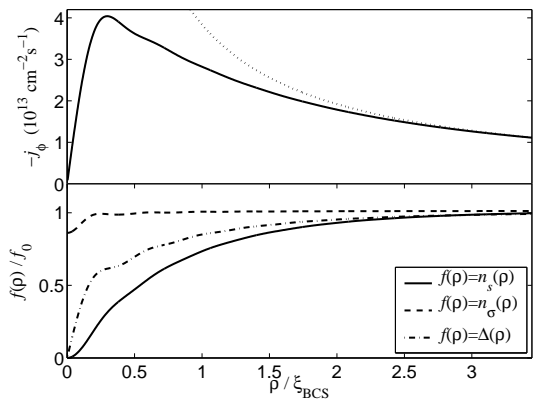,width=7cm,angle=0}
\caption{Upper panel: the current density
${\mathbf{j}}_s(\rho)=j_\phi(\rho){\mathbf{e}}_\phi$ (solid line) 
and its asymptotic form (dotted line).
Lower panel:
the density $n_\sigma(\rho)$, the superfluid density $n_s(\rho)$ 
and the gap function $\Delta(\rho)$ normalized to their theoretical bulk values.
In both panels $\tilde{k}_F|a|=0.59$.}
\label{currentfig}
\end{figure}

To examine the superfluid flow associated
with the vortex giving rise to the angular momentum, we plot 
the current density ${\mathbf{j}}_s(\rho)$ given by 
\begin{equation}
{\mathbf{j}}_s(\rho)=\frac{2\hbar}{mi}\sum_\eta
v_{\eta}^*({\mathbf{r}})\rho^{-1}\partial_\phi
v_{\eta}({\mathbf{r}}){\mathbf{e}}_\phi
\end{equation}
in  Fig.~\ref{currentfig} for $\tilde{k}_F|a|=0.59$.
Because the normal component carries no current by construction, the total 
current density may be written as ${\mathbf{j}}_s(\rho)=2n_s{\mathbf{v}}_s$ 
with the superfluid velocity ${\mathbf{v}}_s={\mathbf{e}}_{\phi}\hbar/2m\rho$ 
and the superfluid (or atom-pair) density $n_s$. We plot $n_s(\rho)$ 
defined in this way in Fig.~\ref{currentfig}. Note that unlike dilute 
interacting Bose gases at zero temperature, the superfluid density in a dilute
Fermi superfluid does not have the same behavior as the order parameter 
$\Delta({\mathbf r})$. As expected, $n_s(\rho)\simeq n_\sigma$ far 
away from the vortex core and $n_s(\rho)\rightarrow 0$ for $\rho\rightarrow0$. 
The dotted line in Fig.~\ref{currentfig} gives the analytical result 
$j_\phi(\rho)=\hbar\tilde{k}_F^3/(6\pi^2m\rho)$, which agrees well with the
numerics away from the vortex core. As a self-consistency check on the 
numerics, the angular momentum per particle along $z$ is found to be
exactly $\hbar/2$, corresponding to one unit of angular momentum $\hbar$ 
per Cooper pair as expected.

As shown in Fig.~\ref{currentfig}, the presence of a vortex in a Fermi system 
does not lead to any significant change in the particle density
$n_\sigma(\rho)$~\cite{deGennes}, in contrast with a dilute Bose gas,
where the density is minuscule on the vortex
line~\cite{Fetter1972}. Direct observation 
of the vortex core (now commonplace for BECs) is therefore not likely. 
The quantized currents, and therefore the presence of superfluidity, can be 
readily detected using at least three approaches, however. One of these is the
collective mode spectrum. When no vortex is present, excitations carrying 
equal and opposite angular momentum along the $z$-axis are degenerate in 
energy. The vortex currents lift this degeneracy since the rotational symmetry 
is removed. The resultant splitting of the surface modes is proportional to 
the angular momentum of the gas~\cite{BruunViverit,BoseDetect}. This technique 
has been used to infer the presence of a vortex in a trapped BEC~\cite{Chevy}. 
A second approach was demonstrated in a recent experiment where the precession 
rate of the scissors 
oscillation mode was used to measure the quantized angular momentum per 
particle with great accuracy~\cite{Foot}. 
A third method is spatially selective 
Bragg scattering; the superfluid currents modify the Bragg momentum 
conservation conditions, giving rise to a strongly anisotropic outcoupled 
atomic beam~\cite{Blakie}.

In conclusion, we have accurately determined the energy of an isolated vortex line
in a dilute superfluid Fermi gas at zero temperature. The results clearly 
indicate that the BCS coherence length sets the scale for the vortex energy
and therefore the critical frequency for vortex stability.

We acknowledge valuable discussions with \O.\ Elgar\o y, C.\ J.\
Pethick, and B.~I.~Schneider.


\begin{references}
\bibitem{Experiments} B.\ DeMarco, S.\ B.\ Papp,  and D.\ S.\ Jin, Phys.\ Rev.\ Lett.\ \textbf{86}, 5409 
(2001); A.\ G.\ Truscott \textit{et al.}, Science \textbf{291}, 2570 (2001);
F.\ Schreck \textit{et al}., Phys.\ Rev.\ A \textbf{64}, 011402 (2001);
S.\ R.\ Granade \textit{et al}., Phys.\ Rev.\ Lett.\ \textbf{88}, 120405 (2002); Z.\ Hadzibabic  \textit{et al}., 
Phys.\ Rev.\ Lett.\ \textbf{88}, 160401 (2002); G.\ Roati \textit{et al}.,
Phys.\ Rev.\ Lett.\ \textbf{89}, 150403 (2002).
\bibitem{StoofBCS} H.\ T.\ C.\ Stoof, M. Houbiers, C.\ A.\ Sackett,
and R.\ G.\ Hulet,  Phys.\ Rev.\ Lett.\ \textbf{76}, 10 (1996).
\bibitem{Fetter} A.\ L.\ Fetter and A.\ A.\ Svidzinsky, J.\
Phys.:Cond.\ Matt.\ \textbf{13}, R135 (2001). 
\bibitem{deGennes} \ P.\ G.\ de Gennes,
\textit{Superconductivity of Metals and Alloys}  
(Addison-Wesley, Reading, MA, 1989).
\bibitem{He3} O.\ Lounasmaa and E.\ Thuneberg, Proc.\ Natl.\ Acad.\ Sci.\ USA~\textbf{96}, 7760 (1999).
\bibitem{ElgaAstro} \O.\ Elgar\o y and F.\ V.\ de Blasio, Astron.\ \&
Astrophys.\ \textbf{370}, 939  
(2001).
\bibitem{Gross} E.\ P.\ Gross, Nuovo Cimento~\textbf{20}, 454 (1961);
L.\ P.\ Pitaevskii,~Zh.\ Eksp.\ Teor.\  
Fiz.~\textbf{40}, 646 (1961) [JETP~{\bf 13}, 451 (1961)]. 
\bibitem{pwavethreshold} B.~DeMarco {\it et al.}, Phys. Rev. Lett.~{\bf 82},
4208 (1999). 
\bibitem{BruunViverit} G.\ M.\ Bruun and L.\ Viverit, Phys.\ Rev.\ A
\textbf{64}, 063606 (2001). 
\bibitem{Induced} We exclude in this paper the effect of induced interactions 
which lowers the magnitude  
of the pairing field. See e.g.  L. P. Gorkov and T. K. Melik-Barkhudarov, 
Sov. Phys. JETP {\bf 13}, 1018 (1961); H. Heiselberg, C. J. Pethick,
H. Smith, and L. Viverit, Phys. Rev. Lett. {\bf 85}, 2418 (2000).
\bibitem{ElgaroyVortex} \O.\ Elgar\o y, e-print: cond-mat/0111440.
\bibitem{Kramer} L.\ Kramer and W.\ Pesch, Z.\ Physik \textbf{269}, 59 (1974).
\bibitem{Gygi} F.\ Gygi and M.\ Schl\"{u}ter, Phys.\ Rev.\ B
\textbf{43}, 7609 (1991).
\bibitem{BruunBCS} G.\ M.\ Bruun, Y.\ Castin, R.\ Dum, and K.\
Burnett, Eur.\ Phys.\ J.\ D \textbf{9}, 433 (1999);  Aurel Bulgac and Yongle 
Yu, Phys.\ Rev\ Lett.~\textbf{88}, 042504 (2002).
\bibitem{JohnThomas} K.\ M.\ O'Hara {\it et al.}, Phys.\ Rev.\ 
Lett.~\textbf{85}, 2092 (2000).
\bibitem{FederDVR}
B.~I. Schneider and D.~L. Feder, Phys. Rev. A {\bf 59},  2232  (1999).
\bibitem{Lemoine}
D. Lemoine, J. Chem. Phys. {\bf 101},  1  (1994).
\bibitem{Fetter1972} A.\ L.\ Fetter, Ann.\ Phys.\ (NY) \textbf{70}, 67 (1972).
\bibitem{BoseDetect} F.\ Zambelli and S.\ Stringari, Phys. Rev. Lett. 
\textbf{81}, 1754 (1998); S.\ Sinha, Phys. Rev. A \textbf{55}, 4325 (1997); 
R.\ J.\ Dodd, K.\ Burnett, M.\ Edwards, and C.\ W.\ Clark, Phys. Rev. 
A~\textbf{56}, 587 (1997); A.\ A.\ Svidzinsky and A.\ L.\ Fetter, Phys. Rev. 
A~\textbf{58}, 3168 (1998).
\bibitem{Chevy} F.\ Chevy, K.\ W.\ Madison, and J.\ Dalibard,  
Phys. Rev. Lett. \textbf{85}, 2223 (2000).
\bibitem{Foot} E.~Hodby {\it et al.}, e-print: cond-mat/0209634.
\bibitem{Blakie} P.~B.~Blakie and R.~J.~Ballagh, Phys. Rev. Lett.~{\bf 86}, 
3930 (2001).
\end{references}
\end{document}